# Negative GMR Effect in current perpendicular-to-plane (Zn, Cr)Te / Cu / Co spin salves

W. G. Wang,[1] C. Ni,[2] L. R. Shah,[1] X.M. Kou[1] and J. Q. Xiao[1]

1) Department of Physics and Astronomy, University of Delaware, Newark, DE 19716, USA

2) Department of Materials Science and Engineering, University of Delaware, Newark, DE 19716,USA

Magnetic and transport properties are explored in the current perpendicular-to-plane (CPP) spin salves with Cr doped wide band gap semiconductor ZnTe as one of the ferromagnetic electrodes. A negative magnetoresistance is observed in these CPP spin valves at low temperature, with a strong temperature dependence. This effect can be explained by the large difference of spin scattering asymmetry coefficients in (Zn,Cr)Te and Cobalt, due to the very different spin polarizations of the two materials as revealed by the DFT calculation.

In normal giant magnetoresistance (GMR) effect, [1-3] the resistance of a spin-valve is low when the magnetizations of two electrodes are in parallel configuration and high when the magnetizations are in antiparallel configuration. In some spin-valves, however, the high resistance state occurs when the magnetizations of two electrodes are parallel to each other. This is the so-called negative ( or inversed) GMR effect. Generally there are two classes of negative GMR effect. In the first class, the negative GMR is caused by different spin scattering asymmetry at two interfaces of the non-magnetic layer. The spin scattering asymmetry coefficient, $\alpha$, is defined by $\alpha = \rho_\downarrow / \rho_\uparrow$, where $\rho_\downarrow$ and $\rho_\uparrow$ are the spin resolved resistivities for spin-down channel and spin-up channel, respectively. Since $\rho_\downarrow$ and $\rho_\uparrow$ are directly proportional to the density of states at the Fermi level, the negative GMR is expected when the spin-valve is consisted by two ferromagnetic electrodes with spin polarizations of different signs. A large difference in spin scattering asymmetry coefficients is often required to observe the negative GMR effect experimentally. For example, the spin polarization is positive for Fe and negative for Co, thus $\alpha_{Fe} < 1$ and $\alpha_{Co} > 1$. But negative GMR was not observed in Fe/Au/Co spin-valves. Only after Fe electrode was doped with V, shifting the spin-up band of Fe upward thus giving rise to a larger positive spin polarization, negative GMR was then observed [4]. Similar negative GMR effect was also observed in $Fe_3O_4$／Au／Fe spin valves.[5]

The second class of negative GMR effect is not due to the intrinsic spin scattering symmetry, instead, it is caused by extrinsic origins like antiferromagnetic (AFM) interaction of magnetic materials in the electrode. For example, negative GMR was observed in CoFeGd/Ag/CoFe spin valves.[6] In the CoFeGd electrode, the total magnetization is in the same direction as Gd moments, which is opposite to the CoFe moments due to the AFM interaction between Gd and CoFe. But the conduction electrons are scattered mostly by the 3d moments of Co and Fe, little affected by the deep 4f moments of Gd. Therefore when the total magnetizations of CoFeGd and CoFe are in parallel configuration, the effective moments in two electrodes contributing to GMR are actually antiparallel to each other, thus resulting in observed negative GMR effect.

Here we present the study of negative GMR effect in spin valves composed of the magnetic semiconductor (Zn,Cr)Te and Co electrodes. (Zn,Cr)Te is an important magnetic semiconductor with high Curie temperature. [7-13] Density of states (DOS) calculation for the two electrodes is performed. It is found the DOS for the majority spins and minority spins at the Fermi level of (Zn,Cr)Te is very much different from that of Co. Unlike Co, in which minority spins have a larger DOS at Fermi level, (Zn,Cr)Te's DOS at Fermi level is dominated by majority spins. It is believed this strong difference of DOS at Fermi level between (Zn,Cr)Te and Co results in the huge difference of spin scattering asymmetries, which leads to the negative GMR effect.

The spin-valves were fabricated by magnetron sputtering with the structure of Si / (Zn,Cr)Te (50 nm )/ Cu ( 1-4 nm) / Co (15 nm) / Cu (70 nm). The Cr concentration in (Zn,Cr)Te is controlled as 10% as described earlier.[14,15] The spacer Cu layer was fabricated as a wedge to located the optimal thickness for the largest MR. After the fabrication of blanket films, the samples were patterned into current-perpendicular-to-plane (CPP) spin-valves and tested in a Physical Property Measurement System (PPMS) at low temperatures.

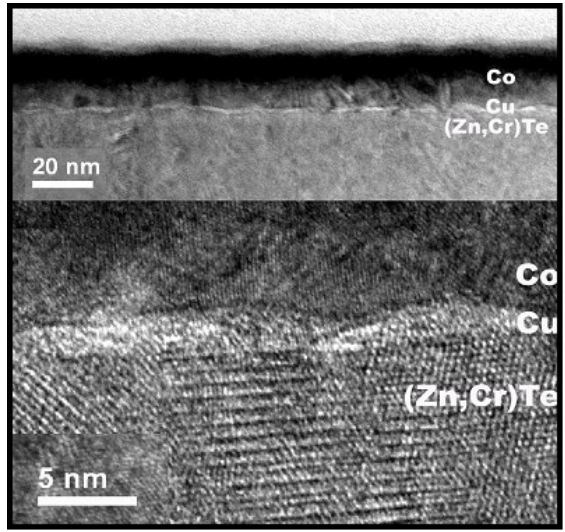

Figure 1  TEM pictures for (Zn,Cr)Te (50 nm )/ Cu ( 1-3 nm) / Co (15 nm) / Cu (70 nm) spin-valves

To investigate the microstructure of the spin-valves, TEM study was performed on the unpatterned films. Figure 1 shows one of the TEM micrographs for the sample. In the low magnification image, we can clearly distinguish the (Zn,Cr)Te, Cu and Co layers. The Cu space layer has a larger roughness due to the thick (Zn,Cr)Te layer underneath, as in the case of (Zn,Cr)Te/$Al_2O_3$/Co MTJs. [16] This indicates larger MR could be achieved upon improving the interface sharpness between (Zn,Cr)Te and Cu layers. Although the Cu spacer layer has large roughness, it is continuous as shown in the high magnification image in Figure 1. (Zn,Cr)Te with the crystal size about 10 nm can also be seen. Interestingly, the Co top electrode grown on Cu has a very good texture.

The magnetic property of the spin-valves was investigated by SQUID. The hysteresis loop measured at 5K is shown in Figure 2.  Two distinct switchings corresponding to  Co and (Zn,Cr)Te layers can be identified, with switching fields matching with that in single layer of Co and (Zn,Cr)Te films. This indicates these two layers are magnetically separated.

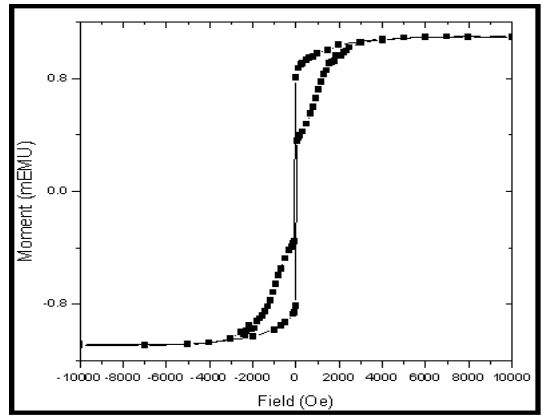

Figure 2    Hysteresis loop for the spin-valve measured at 5 K by SQUID

The negative GMR effect was observed in all samples with Cu thicknesses ranging from 1nm to 3nm at low temperatures as shown in Figure 3. These MR curves generally show large noise due to very small MR ratio of less than one percent. Nevertheless, one can clearly see that the spin-valves are in the low resistances state at antiparallel configuration. The negative MR is about 0.42% with 1.8 nm of Cu, increases to its maximum of 0.45% with 2nm of Cu and starts to decrease with further increasing of the Cu thickness. The noise background can be potentially reduced through thermal annealing, [17] which will be investigated in our future studies.

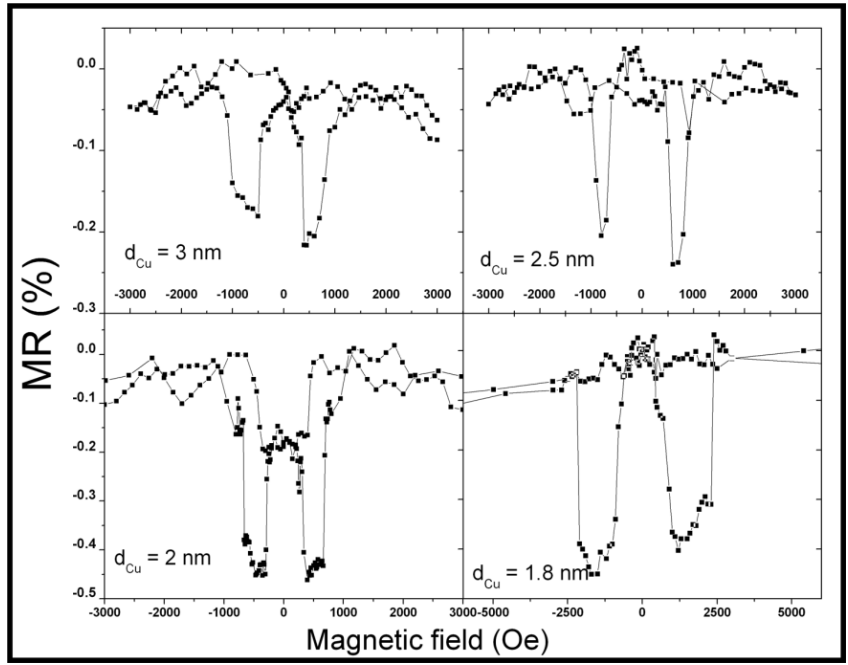

Figure 3    Negative GMR in (Zn,Cr)Te (50 nm )/ Cu ( 1-3 nm) / Co (15 nm) / Cu (70 nm) spin-valves at 10 K.

In order to investigate the origin for the observed negative GMR effect, the density of states for ZnTe, (Zn,Cr)Te and Co were calculated using density functional theory with WIEN 2K package. Two unit cells were used in the calculation of (Zn,Cr)Te, with one Zn atom replace by Cr. This gives the doping concentration of 12.5% since each unit cell contains four Zn atoms. While the actual Cr concentration in the spin-valves is 10%, we believe the calculation performed on 12.5% can still give a good approximation for studying the properties we are interested in. As shown in Figure 4, The undoped ZnTe is nonmagnetic semiconductor, with a band gap about 2.2 eV at zero temperature. After introducing 12.5% Cr into the system, the bands split with unbalanced numbers of spin-up and spin-down electrons. Moreover, the (Zn,Cr)Te in this ideal structure has an 100% positive spin polarization, with only spin up electrons at the Fermi level. This calculation is consistent with the work done by others.[18] On the other hand, as it is well known Co has a very difference DOS distribution at the Fermi level. The Fermi level of Co resides on the tail of the highest energy DOS peak for the spin-up electrons while there is a much larger DOS with spin-down electrons, giving rise to a negative spin polarization. Now the observed negative MR can be explained by the different spin scattering asymmetry coefficients for Co and (Zn,Cr)Te. Obviously, we can see $\alpha_{Co} > 1$ and $\alpha_{ZnCrTe} < 1$ (even equals to zero in the ideal case). This strong difference in $\alpha_{ZnCrTe}$ and $\alpha_{Co}$ results in the negative GMR effect in (Zn,Cr)Te/Cu/Co spin-valves.

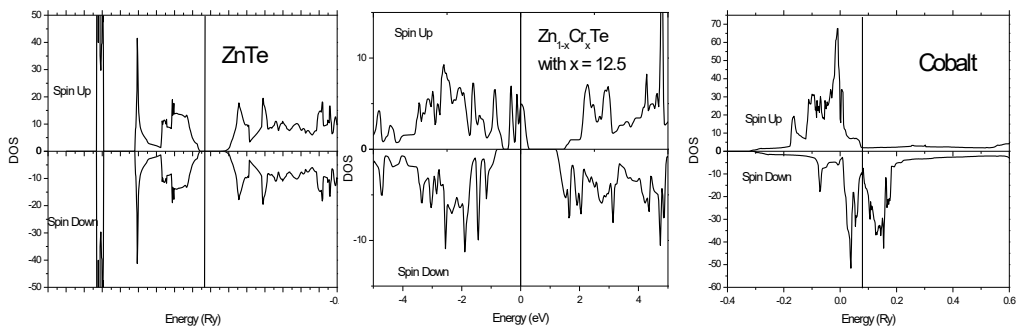

Figure 4    DOS diagrams for ZnTe, (Zn,Cr)Te and Co calculated by DFT method.

Figure 5 shows the temperature dependence of the negative GMR for the spin-valve with 2nm of Cu spacer layer. The negative MR decreases rapidly with increasing temperature as reported in other systems[4,5,19]. This fast decay of MR is possibly due to the rapid increase of spin-flip scattering with temperature, and the non-ideal DOS of polycrystalline (Zn,Cr)Te ( not exactly like "half-metal" as shown Figure 4) thus α is very sensitive to any minor change of Fermi level position[4,19].

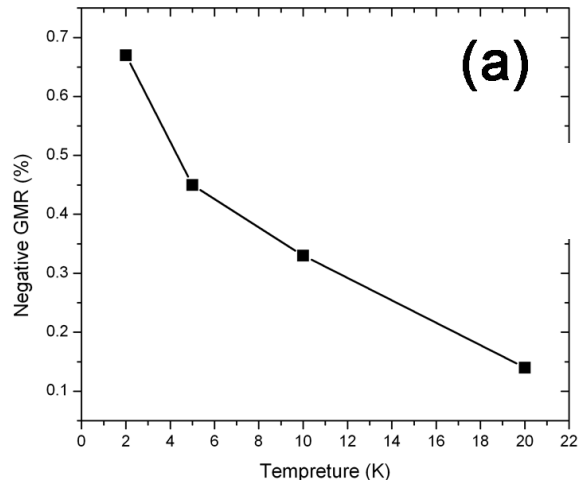

Figure 5    The temperature dependence of the negative GMR effect for the spin-valve with 2nm Cu spacer layer.

It is interesting to compare the negative GMR in (Zn,Cr)Te/Cu/Co spin-valves with the positive TMR in (Zn,Cr)Te/Al$_2$O$_3$/Co tunnel junctions.[16] The only difference in these two systems is the spacer layer. The absence of negative TMR in the Co based junction has been investigated for a long time, as described in the review article of Tsymbal *et al*.[20] It is generally believed that the tunneling spin polarization for Co electrode in contact with Al$_2$O$_3$ is positive, although the total DOS for Co is negative. It is because that the total DOS in Co is dominated by *d* band electrons, which are more localized; whereas the *sp* band electrons with positive spin polarization determines the sign of TMR to be positive. The observation of the negative TMR in the Co junctions is possible with very thin Al$_2$O$_3$ barrier. Although the *d* band and *s* band have different spin polarization in Co, the characteristics of *d* band still impacts the tunneling spin polarization through formation of interfacial bonding states.[20] The *d* characteristics can be coupled to the s states inside the barrier, which then quickly decays. Therefore, observation of negative TMR in MTJs like Co/Al$_2$O$_3$/Fe may require very thin barrier (<1nm). Practically it is very hard to fabricate a pinhole-free barrier with that thickness. While in the spin-valve structure, the scattering of the conducting electrodes by the *d* band of Co is more easily coupled to the other electrode through the metallic spacer layer, thus a negative GMR is detected in (Zn,Cr)Te/Cu/Co spin-valves.

To conclude, a negative GMR is observed in the in (Zn, Cr)Te / Cu / Co Spin Valves. Magnetoresistance measurement at low temperature shows consistent behavior of negative GMR over different Cu spacer thickness. This effect can be explained by the large difference of spin scattering asymmetry coefficients in Co and (Zn,Cr)Te, due to the very different spin polarizations of the two materials as revealed by the DFT calculation.

## ACKNOWLEDGEMENT

This work was supported by DOE DE-FG02-07ER46374 and NSF Grant No. DMR0827249.